\documentclass[%
 reprint,
 amsmath,amssymb,
 aps,
 prl
]{revtex4-1}

\usepackage{graphicx}
\usepackage{dcolumn}
\usepackage{bm}

\usepackage[utf8]{inputenc}
\usepackage{amsmath}
\usepackage{amssymb}
\usepackage{epstopdf, epsfig}

\usepackage{color}
\usepackage{amsmath}
\usepackage{amsfonts}
\usepackage{amssymb}
\usepackage{amsbsy}
\usepackage{extarrows}
\usepackage{tikz}
\usepackage{hyperref}
\usepackage{pgfplots}
\usepackage{ifthen}

\usepackage{tensor}

\newcommand{\ie}{i.\,e.}%
\newcommand{\wrt}{w.\,r.\,t.}%

\newcommand{\formComma}{\,\text{,}}
\newcommand{\formPeriod}{\,\text{.}}

\newcommand{\pressure}{p}

\newcommand{\tr}{\operatorname{tr}}

\newcommand{\vbc}{v}%
\newcommand{\vb}{\mathbf{\vbc}}%
\newcommand{\Vb}{\mathbf{V}}
\newcommand{\Xb}{\mathbf{x}}
\newcommand{\Qb}{\mathbf{Q}}
\newcommand{\Thetab}{\mathbf{\Theta}}
\newcommand{\Phib}{\mathbf{\Phi}}
\newcommand{\surf}{\mathcal{S}}
\newcommand{\gaussianCurvature}{K}
\newcommand{\Grad}{\operatorname{grad}}
\newcommand{\Div}{\operatorname{div}}%
\newcommand{\Rot}{\operatorname{rot}}%
\newcommand{\BigRot}{\operatorname{rot}}%
\newcommand{\DivSurf}{\Div_{\surf}}%
\newcommand{\GradSurf}{\Grad_{\surf}}
\newcommand{\RotSurf}{\Rot_{\surf}}
\newcommand{\BigRotSurf}{\BigRot_{\surf}}

\newcommand{\vecLaplace}{\boldsymbol{\Delta}}
\newcommand{\laplaceDeRham}{\vecLaplace^{\textup{dR}}}

\newcommand{\laplaceBochner}{\vecLaplace_{\textup{B}}}
\newcommand{\LaplaceDeRham}{Laplace-deRham }

\newcommand{\levicivitaPlain}{\boldsymbol{\nabla}}
\newcommand{\levicivita}[2]{\levicivitaPlain_{#1}{#2}}
\renewcommand{\Re}{\text{Re}}

\newcommand{\killingVF}{\vb_{\text{K}}}

\newcommand{\harmspace}{{\mathcal{H}}}
\newcommand{\divfreespace}{{\operatorname{Ker}\DivSurf}}

\begin{document}





\begin{abstract}
\end{abstract}

\pacs{Valid PACS appear here}
\maketitle


\section{Comment on ``Geometrical Control of Active Turbulence in Curved Topographies"}

In a recent letter \cite{Pearceetal_prl_2019} Pearce et al. investigate the turbulent dynamics of a two-dimensional active nematic liquid crystal which is constrained to a curved surface. 
The underlying model combines an incompressible surface Navier-Stokes equation with friction and active forcing with a surface Landau-de Gennes model for nematic liquid crystals. 
To solve the surface Navier-Stokes equation a vorticity-stream function approach is considered.
The approach is based on a decomposition of the velocity field in divergence- and curl-free parts. 
While appropriate on simply connected surfaces, this is not sufficient on non-simply connected surfaces, such as the considered torus. 
As a consequence of the topology also non-trivial harmonic parts, velocity fields which are divergence- and curl-free, exist, which are not represented by the vorticity-stream function approach, see \cite{Nitschkeetal_book_2017} for an example. We here explain the underlying situation and provide details and examples in the Supplementary Information (SI).

Consider the surface Navier-Stokes equation with friction and forcing terms, eq. (1a) of \cite{Pearceetal_prl_2019}. 
Without the friction and forcing terms, with the exception of some special initial conditions for which the solution converges to zero, any solution of the surface Navier-Stokes equation on a torus converges to a stationary Killing vector field, which contains non-trivial harmonic parts, see \cite{Nitschkeetal_book_2017,Reutheretal_PF_2018}. 
Clearly, with the friction and without the forcing term, the velocity converges to zero.
In contrast to the argumentation in \cite{Pearceetal_prl_2019} this means that the harmonic and the anti-harmonic part vanish identically over time.
Moreover, considering also the forcing term introduces additional harmonic parts to the velocity. 
There is no reason to assume that these solution components are negligible. In fact neglecting them changes the velocity qualitatively in an nonphysical manner. In the SI we consider an example which takes friction and forcing via a constructed Q tensor in the sense of eq. (1a) of \cite{Pearceetal_prl_2019} into account and demonstarte that the harmonic parts are not negligible. We further show that the argumentation following eq. (S43) in the SI of \cite{Pearceetal_prl_2019} is not correct.

To solve the active nematic liquid crystal model considered in \cite{Pearceetal_prl_2019} on general curved topographies thus requires an approach which also handles the harmonic parts of the velocity field and directly acts on the velocity and pressure variables. 
Numerical approaches can be found in \cite{Nitschkeetal_book_2017} using discrete exterior calculus (DEC) and in \cite{Reutheretal_PF_2018,Nitschkeetal_PRF_2019} using surface finite elements for each component of an extended velocity field in the embedding space and a penalization of the normal component. 
General approaches how to solve surface vector- and tensor-valued partial differential equations can be found in \cite{Nesteretal_JCP_2019}.\\

\noindent Ingo Nitschke, Sebastian Reuther \\
\hspace*{0.5cm}Institute of Scientific Computing, TU Dresden \\
\hspace*{0.5cm}01062 Dresden, Germany \\

\noindent Axel Voigt \\
\hspace*{0.5cm}Institute of Scientific Computing, TU Dresden and \\
\hspace*{0.5cm}Dresden Center for Computational Materials Science \\
\hspace*{0.5cm}(DCMS) and \\
\hspace*{0.5cm}Center for Systems Biology Dresden (CSBD) and \\
\hspace*{0.5cm}Cluster of Excellence Physics of Life (PoL) \\
\hspace*{0.5cm}01062 Dresden, Germany

\bibliography{comment}

\newpage

\section{Supplementary Information}

We follow the notation of \cite{Pearceetal_prl_2019} but use a more compact vector notation for the differential operators, see \cite{Nitschkeetal_book_2017}.
Let $\divfreespace$ be the space of divergence free vector fields on the surface $\surf$.
This space can be decomposed \wrt\ the $L_2$ inner product into parts with and without curl-components by using the Hodge theorem, \ie
\begin{align*}
    \divfreespace &= \RotSurf \mathcal{C}^\infty(\surf) \oplus_{L_2} \harmspace(\surf),
\end{align*}
where the last considers the harmonic parts. Here, we explicitly note that this decomposition does not hold in the sense of the local inner product, but in the sense of the $L_2$ inner product.
In the following we restrict to the surface of the torus with radii ratio $\xi := a/b$ with major and minor radius $a$ and $b$, respectively, and denote the two basis vector fields by $\partial_\theta\Xb$ and $\partial_\phi\Xb$ as well as the two basis harmonic vector fields by $\Thetab$ and $\Phib$, as considered in \cite{Pearceetal_prl_2019}.
It is thereby noted that these harmonic vector fields are locally orthogonal to each other \wrt\ the local inner product. 
Thus, an orthogonal projection of a vector fields into the space of harmonic vector fields $\Pi_\harmspace: T^1\surf \rightarrow \harmspace(\surf)$ can be considered using basic orthogonalization techniques, \ie
\begin{align*}
    \Pi_{\harmspace} \vb &= \frac{\langle \vb , \Thetab \rangle_{L_2}}{\|\Thetab\|_{L_2}^2}\Thetab + \frac{\langle \vb , \Phib \rangle_{L_2}}{\|\Phib\|_{L_2}^2}\Phib \formComma
\end{align*}
where $T^1\surf$ denotes the space of vector fields on $\surf$.
Using this, the anti-harmonic projection $\Pi^\bot_\harmspace: \divfreespace \rightarrow \RotSurf \mathcal{C}^\infty(\surf)$ is defined by 
\begin{align*}
    \Pi^\bot_{\harmspace} \vb &= \vb - \Pi_{\harmspace} \vb \formPeriod
\end{align*}

We recall eq. (1a) of \cite{Pearceetal_prl_2019} in vector notation, \ie
\begin{align}
    \label{eq:sns}
    \rho\frac{D\vb}{Dt} &= \eta\left(\laplaceBochner\vb + \gaussianCurvature\vb\right) - \GradSurf\pressure - \zeta\vb + \alpha\DivSurf\Qb
\end{align}
where $\vb$ denotes the velocity, $\frac{D}{Dt}$ the covariant material time derivative, $\laplaceBochner$ the Bochner Laplacian, $\GradSurf$ the surface gradient, $\DivSurf$ the surface diverence,  $\gaussianCurvature$ the Gaussian curvature, $\pressure$ the surface pressure, $\Qb$ a Q-tensor, $\rho$ the material density, $\eta$ the dynamic viscosity, $\zeta$ the friction coefficient and $\alpha$ the strength of the external forcing. 

In the following we mainly focus on three different configurations of eq. \eqref{eq:sns}.
Firstly, no external contributions are considered, \ie\ $\zeta=0$ and $\alpha=0$. 
Secondly, we investigate damped flow, \ie\ $\zeta\neq0$ and $\alpha=0$, and show that there is a strong influence of the harmonic contributions. 
This is the same setup as in the Supplementary Information of \cite{Pearceetal_prl_2019}, where it is argued that harmonic contributions are negligible. 
Thirdly, the external contributions are considered, \ie\ $\zeta\neq0$ and $\alpha\neq0$.
In this case we demonstrate on an example that the harmonic parts do not vanish over time. 

Additionally, we show that the argumentation in the Supplementary Information of \cite{Pearceetal_prl_2019} is based on a wrong assumption and therefore contradicts with the following argumentation in \cite{Pearceetal_prl_2019}. 

\subsection{Flow with no external contributions}

Let $\zeta=0$ and $\alpha=0$. 
The solution of eq. \eqref{eq:sns} converges to a Killing vector field for $t\rightarrow\infty$ (except for some special initial data with symmetric behavior which implies a vanishing solution), see \cite{Nitschkeetal_book_2017}.
These vector fields exist on rotationally symmetric surfaces and are characterized by a vanishing deformation tensor. 
In the present case of a torus, the Killing vector field -- denoted by $\killingVF$ -- can be determined by $\killingVF = \alpha_K\partial_\phi\Xb$ with a constant $\alpha_K$ and describes a rigidly rotating torus. 
It can be easily verified that $\Pi_{\harmspace}\killingVF\neq 0$ and thus, non-trivial Killing vector fields contain non-trivial harmonic parts, which cannot vanish asymptotically in general.
Cancelling out the harmonic part for a rigidly rotating torus would magically generate internal friction and the velocity dissipates to zero without any physical reason.

\subsection{Damped flow}

\begin{figure}[ht!]
    \centering
    \begin{minipage}{0.49\textwidth}
%
%
\begin{tikzpicture}

\begin{axis}[%
width=0.8\textwidth, 
height=0.3\textwidth, 
at={(0,0)}, 
scale only axis,
separate axis lines,
every outer x axis line/.append style={black},
every x tick label/.append style={font=\color{black}},
xmin=0,
xmax=40,
xtick={  0,  20,  40,  60},
xlabel={$\mbox{Time}$},
every outer y axis line/.append style={black},
every y tick label/.append style={font=\color{black}},
ymode=log,
ymin=1e-02,
ymax=1,
yminorticks=true,
ylabel={$\mbox{Kinetic energy}$},
x label style={font=\footnotesize,at={(axis description cs:0.5,0.075)}}, 
y label style={font=\footnotesize,at={(axis description cs:0.025,0.5)}}, 
yticklabel style = {font=\footnotesize}, 
xticklabel style = {font=\footnotesize}, 
axis background/.style={fill=white},
legend style={font=\scriptsize,at={(0.99,0.99)},anchor=north east,legend cell align=left,align=left,draw=black} 
]
\addplot [color=black,solid,line width=2.5pt]
  table[row sep=crcr]{%
0	1\\
1	0.905286954675\\
2	0.819544470304\\
3	0.741922917743\\
4	0.671653138807\\
5	0.608038824628\\
6	0.550449615872\\
7	0.498314856454\\
8	0.451117938869\\
9	0.408391185078\\
10	0.369711212255\\
11	0.334694737452\\
12	0.302994779613\\
13	0.274297221318\\
14	0.248317696163\\
15	0.224798770951\\
16	0.203507394769\\
17	0.184232589664\\
18	0.166783360049\\
19	0.150986800109\\
20	0.136686380467\\
21	0.123740397119\\
22	0.112020567278\\
23	0.101410758212\\
24	0.0918058364729\\
25	0.0831106261219\\
26	0.075238965623\\
27	0.0681128540618\\
28	0.0616616782278\\
29	0.055821512903\\
30	0.0505344874213\\
31	0.0457482122237\\
32	0.0414152597258\\
33	0.0374926943542\\
34	0.0339416470945\\
35	0.0307269303348\\
36	0.0278166891893\\
37	0.0251820858453\\
38	0.0227970138073\\
39	0.0206378392053\\
40	0.0186831666052\\
};
\addlegendentry{$\mbox{full}$};

\addplot [color=red,solid,line width=1.5pt]
  table[row sep=crcr]{%
0	0.793622490418\\
1	0.718456087512\\
2	0.650408923532\\
3	0.588806713677\\
4	0.533039036717\\
5	0.482553286272\\
6	0.436849194998\\
7	0.395473877392\\
8	0.358017342117\\
9	0.324108429366\\
10	0.293411133005\\
11	0.265621271066\\
12	0.24046347158\\
13	0.217688443898\\
14	0.197070508444\\
15	0.178405360445\\
16	0.161508045455\\
17	0.146211126626\\
18	0.132363025563\\
19	0.119826520323\\
20	0.108477385673\\
21	0.0982031621266\\
22	0.0889020415811\\
23	0.0804818584873\\
24	0.0728591765765\\
25	0.0659584620831\\
26	0.0597113352742\\
27	0.05405589287\\
28	0.0489360946385\\
29	0.044301208089\\
30	0.0401053057593\\
31	0.0363068101171\\
32	0.0328680815649\\
33	0.0297550454659\\
34	0.026936854496\\
35	0.0243855829752\\
36	0.0220759501496\\
37	0.0199850696825\\
38	0.0180922228718\\
39	0.0163786533469\\
40	0.0148273812101\\
};
\addlegendentry{$\mbox{harmonic}$};

\addplot [color=blue,solid,line width=1.5pt]
  table[row sep=crcr]{%
0	0.608410505089\\
1	0.550786093344\\
2	0.498619465121\\
3	0.451393697121\\
4	0.408640825426\\
5	0.369937208405\\
6	0.334899328818\\
7	0.303179993509\\
8	0.274464893042\\
9	0.248469487187\\
10	0.224936185385\\
11	0.203631794263\\
12	0.184345206904\\
13	0.166885310967\\
14	0.151079094945\\
15	0.136769933778\\
16	0.123816036841\\
17	0.112089042932\\
18	0.101472748328\\
19	0.0918619553163\\
20	0.0831614297788\\
21	0.0752849575108\\
22	0.0681544899178\\
23	0.0616993706251\\
24	0.0558556353386\\
25	0.0505653780171\\
26	0.0457761770771\\
27	0.0414405759427\\
28	0.0375156127952\\
29	0.0339623948601\\
30	0.0307457130164\\
31	0.0278336929059\\
32	0.0251974790881\\
33	0.0228109491092\\
34	0.0206504546523\\
35	0.0186945872048\\
36	0.0169239659196\\
37	0.0153210455683\\
38	0.013869942685\\
39	0.0125562781748\\
40	0.0113670348309\\
};
\addlegendentry{$\mbox{anti-harmonic}$};

\end{axis}
\end{tikzpicture}%
    \end{minipage}
    \caption{Kinetic energy of the Killing solution as well as its harmonic and anti-harmonic parts for $a=2$, $b=1$, $\rho=1$, $\eta=0.1$, $\zeta=0.1$ and $\alpha=0$. The initial Killing vector field $\killingVF(0)$ is rescaled, such that its $L_2$-norm is normalized.}
    \label{fig:killingsemilogy}
\end{figure}
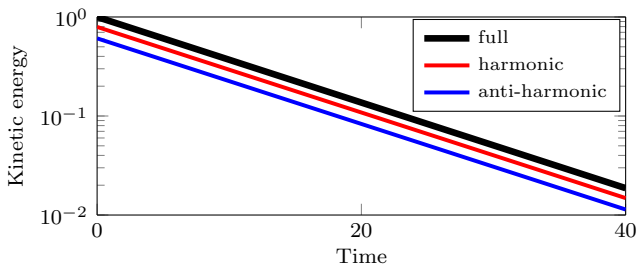

If we consider external friction, \ie\ $\zeta\neq0$, and no additional forces, \ie\ $\alpha=0$, then we indeed agree that the harmonic part of the solution vanishes exponentially.
But the anti-harmonic part do the same and we do not see any reasons to only highlight this behavior for the harmonic part. 
In particular, it can be verified that the damped Killing solution $\killingVF(t) = \killingVF(0)\exp{(-\zeta/\rho t)}$ solves eq. \eqref{eq:sns} with the Killing vector field from above as initial condition.
Thus, the harmonic part $\Pi_{\harmspace}\killingVF$ and the anti-harmonic part 
$\Pi_{\harmspace}^\bot \killingVF$
vanish exponentially with the same order (in time). 
\autoref{fig:killingsemilogy} shows this behavior, where the kinetic energy of the full damped Killing vector field as well as its harmonic and anti-harmonic part is shown. 
The harmonic contribution is even larger than the anti-harmonic part. 
Moreover, the ratio between these parts can be determined by 
\begin{align*}
    r_{\mathcal{H}} &:= \frac{\|\Pi_\harmspace\killingVF(t)\|_{L_2}}{\|\Pi_\harmspace^\bot\killingVF(t)\|_{L_2}}
          = \left(\frac{2\xi\left(\xi^2-1\right)^{1/2}}{3 + 2\xi^2 - 2\xi\left(\xi^2-1\right)^{1/2}}\right)^{1/2}
\end{align*}
which only depends on the radii ratio $\xi$ of the torus and is monotonically increasing for $\xi>1$.
Furthermore, $\xi>((7+2\cdot19^{1/2})/6)^{1/2}\approx 1.619$ yields $r_{\mathcal{H}}>1$ and thus 
larger harmonic parts.

\subsection{Damped and forced flow}

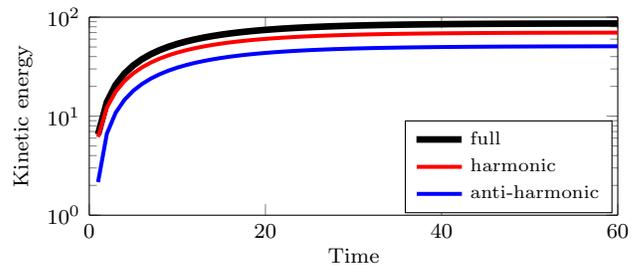
\begin{figure}[ht!]
    \centering
    \begin{minipage}{0.49\textwidth}
%
%
\begin{tikzpicture}

\begin{axis}[%
width=0.8\textwidth, 
height=0.3\textwidth, 
at={(0,0)}, 
scale only axis,
separate axis lines,
every outer x axis line/.append style={black},
every x tick label/.append style={font=\color{black}},
xmin=0,
xmax=60,
xtick={  0,  20,  40,  60},
xlabel={$\mbox{Time}$},
every outer y axis line/.append style={black},
every y tick label/.append style={font=\color{black}},
ymode=log,
ymin=1e-0,
ymax=1e2,
yminorticks=true,
ylabel={$\mbox{Kinetic energy}$},
x label style={font=\footnotesize,at={(axis description cs:0.5,0.075)}}, 
y label style={font=\footnotesize,at={(axis description cs:0.05,0.5)}}, 
yticklabel style = {font=\footnotesize}, 
xticklabel style = {font=\footnotesize}, 
axis background/.style={fill=white},
legend style={font=\scriptsize,at={(0.99,0.01)},anchor=south east,legend cell align=left,align=left,draw=black} 
]
\addplot [color=black,solid,line width=2.5pt]
  table[row sep=crcr]{%
0	0\\
1	6.5303420167\\
2	13.8925684227\\
3	20.7278598438\\
4	26.9436326267\\
5	32.5799616021\\
6	37.6866155364\\
7	42.3117723107\\
8	46.5001121429\\
9	50.2925265856\\
10	53.7262436764\\
11	56.8350740445\\
12	59.6496863914\\
13	62.1978801822\\
14	64.5048437233\\
15	66.5933935696\\
16	68.4841943974\\
17	70.1959598422\\
18	71.7456353744\\
19	73.1485645045\\
20	74.4186396567\\
21	75.5684390176\\
22	76.6093505994\\
23	77.551684672\\
24	78.4047756272\\
25	79.1770742522\\
26	79.8762313041\\
27	80.509173198\\
28	81.0821705473\\
29	81.6009002278\\
30	82.070501575\\
31	82.495627268\\
32	82.8804893993\\
33	83.2289011873\\
34	83.5443147412\\
35	83.829855252\\
36	84.0883519478\\
37	84.3223661185\\
38	84.5342164881\\
39	84.726002184\\
40	84.8996235332\\
41	85.056800888\\
42	85.1990916697\\
43	85.3279057992\\
44	85.4445196653\\
45	85.5500887709\\
46	85.6456591817\\
47	85.7321778905\\
48	85.8105022001\\
49	85.8814082176\\
50	85.9455985444\\
51	86.0037092379\\
52	86.0563161135\\
53	86.1039404504\\
54	86.1470541566\\
55	86.186084445\\
56	86.2214180661\\
57	86.2534051407\\
58	86.282362629\\
59	86.308577471\\
60	86.3323094301\\
};
\addlegendentry{$\mbox{full}$};

\addplot [color=red,solid,line width=1.5pt]
  table[row sep=crcr]{%
0	0\\
1	6.16303802366\\
2	12.1942530363\\
3	17.6674479811\\
4	22.6226484893\\
5	27.1085382903\\
6	31.1695561432\\
7	34.8459426379\\
8	38.1741273723\\
9	41.187089595\\
10	43.9146849902\\
11	46.3839415191\\
12	48.6193272425\\
13	50.6429927765\\
14	52.474990785\\
15	54.1334746832\\
16	55.6348785207\\
17	56.9940798285\\
18	58.2245470413\\
19	59.3384729571\\
20	60.3468955572\\
21	61.2598073819\\
22	62.0862545475\\
23	62.8344263853\\
24	63.5117365899\\
25	64.1248966824\\
26	64.6799825153\\
27	65.1824944785\\
28	65.6374120034\\
29	66.0492429041\\
30	66.422068046\\
31	66.7595817834\\
32	67.0651285669\\
33	67.341736084\\
34	67.5921452608\\
35	67.8188374219\\
36	68.0240588781\\
37	68.2098431852\\
38	68.3780312948\\
39	68.5302897963\\
40	68.6681274315\\
41	68.7929100445\\
42	68.9058741162\\
43	69.0081390167\\
44	69.100718097\\
45	69.1845287307\\
46	69.260401404\\
47	69.3290879454\\
48	69.3912689753\\
49	69.4475606505\\
50	69.4985207697\\
51	69.5446543008\\
52	69.5864183847\\
53	69.624226865\\
54	69.6584543891\\
55	69.68944012\\
56	69.7174910981\\
57	69.7428852826\\
58	69.7658743065\\
59	69.78668597\\
60	69.8055264974\\
};
\addlegendentry{$\mbox{harmonic}$};

\addplot [color=blue,solid,line width=1.5pt]
  table[row sep=crcr]{%
0	0\\
1	2.1592427316\\
2	6.65609872728\\
3	10.8399933367\\
4	14.6347229032\\
5	18.0715536066\\
6	21.183478478\\
7	24.000965767\\
8	26.5517688426\\
9	28.8610790003\\
10	30.951731806\\
11	32.8444152148\\
12	34.5578660378\\
13	36.1090512448\\
14	37.5133337078\\
15	38.7846230591\\
16	39.9355126954\\
17	40.9774040499\\
18	41.9206192358\\
19	42.7745031086\\
20	43.5475157152\\
21	44.2473160212\\
22	44.8808377319\\
23	45.4543579461\\
24	45.9735593193\\
25	46.4435863452\\
26	46.8690963127\\
27	47.2543054398\\
28	47.6030306406\\
29	47.9187273389\\
30	48.2045237013\\
31	48.4632516289\\
32	48.6974748152\\
33	48.9095141465\\
34	49.1014706971\\
35	49.2752465474\\
36	49.4325636301\\
37	49.5749807916\\
38	49.7039092376\\
39	49.8206265168\\
40	49.9262891785\\
41	50.0219442335\\
42	50.1085395279\\
43	50.1869331354\\
44	50.2579018594\\
45	50.3221489309\\
46	50.3803109758\\
47	50.432964324\\
48	50.4806307193\\
49	50.5237824903\\
50	50.5628472298\\
51	50.5982120322\\
52	50.6302273292\\
53	50.6592103623\\
54	50.6854483258\\
55	50.7092012135\\
56	50.730704394\\
57	50.7501709439\\
58	50.7677937584\\
59	50.7837474631\\
60	50.7981901444\\
};
\addlegendentry{$\mbox{anti-harmonic}$};

\end{axis}
\end{tikzpicture}%
    \end{minipage}
    \caption{Kinetic energy of the full solution as well as its harmonic and anti-harmonic parts for the damped and forced setting with $a=2$, $b=1$, $\rho=1$, $\eta=0.1$, $\zeta=0.1$ and $\alpha=1$. As initial condition the trivial solution is used.}
    \label{fig:withforcelong}
\end{figure}

Consider the full equation \eqref{eq:sns} with $\zeta\neq0$ and $\alpha\neq 0$. 
The force $\DivSurf\Qb$ feeds harmonic parts into the solution, since $\DivSurf\Qb$ has harmonic contributions itself in general. 
To highlight this behavior, we construct an appropriate Q-tensor. 
Generally, a Q-tensor can be obtained by $\Qb = \GradSurf\Vb$, where $\Vb\in\harmspace(\surf)$. This holds as
$\tr\GradSurf\Vb = \DivSurf\Vb = 0$ and $\langle\GradSurf\Vb, \boldsymbol{\varepsilon}\rangle = -\RotSurf\Vb=0$ with the Levi-Civita tensor $\boldsymbol{\varepsilon}$.
Thus, $\GradSurf\Vb$ is a Q-tensor and we obtain $\DivSurf\Qb = \gaussianCurvature\Vb$.
For the latter identity we used the Weizenböck identity $\laplaceBochner\vb = \laplaceDeRham\vb + \gaussianCurvature\vb$ with the \LaplaceDeRham operator $\laplaceDeRham$, which can be obtained by $\laplaceDeRham\vb = -\RotSurf\RotSurf\vb - \GradSurf\DivSurf\vb$, see \cite{Nitschkeetal_book_2017}.
Consider the special choice $\Qb=-b^2(\xi^2-1)\GradSurf\Phib$. 
Thus, we obtain $\DivSurf\Qb = -b^2(\xi^2-1)\gaussianCurvature\Phib$ and the resulting harmonic force is given by $\Pi_\harmspace\DivSurf\Qb = \Phib$.
This means we constantly feed the system with the harmonic field $\alpha\Phib$ such that the harmonic part of the solution cannot vanish.
In this situation we expect a balance of internal and external friction with the applied force, such that the solution converges to a stationary non-trivial vector field for $t\rightarrow\infty$. 
We further expect that the reached steady state solution does not have negligible harmonic contributions. 
To demonstrate this we discretize eq. \eqref{eq:sns} in time by using an implicit Euler scheme and in space by using a symmetric ansatz and finite differences in poloidal direction. 
\autoref{fig:withforcelong} shows the kinetic energy of the full solution as well as the harmonic and anti-harmonic components. 
Indeed the harmonic contributions are not negligible and are even larger than the anti-harmonic parts in this simple example.
The harmonic-anti-harmonic ratio $r_{\mathcal{H}}|_{t=60} \approx 1.374$ in the steady state regime also reflects this behavior.

\subsection{Wrong assumption}
In the Supplementary Information of \cite{Pearceetal_prl_2019} the authors restrict to the case $\zeta\neq0$, $\alpha=0$ and argue that the harmonic solution components decay to zero due to the friction term in the surface Navier-Stokes equation. 
As discussed above, this is also true for the anti-harmonic part. 
However, the argumentation in \cite{Pearceetal_prl_2019} is based on eq. (S43) of  \cite{Pearceetal_prl_2019}.
In the following we will show that this equation is not correct.
Let $\Vb\in\harmspace(\surf)$ be a harmonic vector field, \ie\ a linear combination of the basis harmonic vector fields $\Thetab$ and $\Phib$. 
Eq. (S43) of  \cite{Pearceetal_prl_2019} reads in the present notation
\begin{align*}
	\DivSurf\left( \Vb\otimes\Vb \right) &= \frac{1}{\Re}\DivSurf(\GradSurf\Vb + {\GradSurf\Vb}^T) \notag \\ 
	&\qquad 
	- \GradSurf\pressure^h \formPeriod
\end{align*}
With $\DivSurf\left( \Vb\otimes\Vb \right) = \levicivita{\Vb}{\Vb} = \frac{1}{2}\GradSurf\left\langle\Vb,\Vb\right\rangle$ and $\DivSurf(\GradSurf\Vb + {\GradSurf\Vb}^T) = - \laplaceDeRham\Vb + 2\gaussianCurvature\Vb = 2\gaussianCurvature\Vb$, this reduces to 
\begin{align}
	\label{eq:s43:new}
	\GradSurf\pressure^h + \frac{1}{2}\GradSurf\left\langle\Vb,\Vb\right\rangle &= \frac{2}{\Re}\gaussianCurvature\Vb \formPeriod
\end{align}
Taking the curl of eq. \eqref{eq:s43:new} results in 
\begin{align}
	\label{eq:s43:new:2}
	\left\langle\BigRotSurf\gaussianCurvature,\Vb\right\rangle &= 0,
\end{align}
which cannot be true in general. On the considered torus $\BigRotSurf\gaussianCurvature$ points along the toroidal direction, whereas  $\Vb$ is harmonic and points in toroidal and poloidal direction, generally. As a consequence, eq. (S43) of  \cite{Pearceetal_prl_2019} does not hold, which also contradicts the following argumentation in \cite{Pearceetal_prl_2019}.

\begin{acknowledgments}
\end{acknowledgments}


\end{document}